# Ground-state degeneracies leave recognizable topological scars in the one-particle density

Roi Baer

*Fritz Haber Center for Molecular Dynamics, Institute of Chemistry, the Hebrew University of Jerusalem, Jerusalem 91904 Israel.*

In Kohn-Sham density functional theory (KS-DFT) a fictitious system of non-interacting particles is constructed having the same ground-state (GS) density as the physical system of interest. A fundamental open question in DFT concerns the ability of an exact KS calculation to spot and characterize the GS degeneracies in the physical system. In this article we provide theoretical evidence suggesting that the GS density, as a function of position on a 2D manifold of parameters affecting the external potential, is "topologically scarred" in a distinct way by degeneracies. These scars are sufficiently detailed to enable determination of the positions of degeneracies and even the associated Berry phases. We conclude that an exact KS calculation can spot and characterize the degeneracies of the physical system.

Electronic degeneracies in molecular systems have an important role in many photochemical and photophysical processes. Degeneracies induce nonadiabatic transitions [1] but may also affect adiabatic dynamics, due to their geometric phase effects.[2] As a consequence, considerable efforts go into identification and location of degeneracies on the manifold of nuclear coordinates. First principles approaches for these tasks use correlated electronic wave functions [3] which quickly become computationally intensive as the size of systems grows. Because of lack of theoretical understanding and practical difficulties in applications, Kohn-Sham (KS) density functional (DFT) methods [4] are not considered appropriate for degeneracies, in spite of their great success in other aspects of electronic structure.[5]

KS-DFT relies on the mapping of an "interacting" electron system onto a "non-interacting", the KS system, , having the same GS density. Most references to degeneracies by DFT researchers refer to existence, uniqueness and differentiability of the mapping.[6] A rather unique study of degeneracies concluded that degeneracies, while rare in the potential manifold are abundant in the "density manifold" [7]: if a GS density is chosen at random, there is no way to determine a priori whether it corresponds to non-degenerate, doubly degenerate,... GSs. This explains the practical difficulty for application of DFT to degeneracies. However, one should note that a physical electronic density (of some molecule, for example) is not "a random density". It is pure-state v-representable (PVR), meaning that the density is derivable from a single wave function, as opposed to the more general an ensemble density which is a weighted sum of pure densities derived from degenerate states. If the physical potential is varied by some parameters on a manifold we obtain corresponding "physical" PVR densities.

Our study here concentrates on these special but relevant density degeneracies, which may be more tractable than the general case discussed in ref.[7] We show here that such densities carry the original degeneracy information in the form of topological non-analyticities we call "scars". A related issue is how (if at all) the Longuet-Higgins sign-change or Berry phase [8] imprinted on the density. Since density is derivable from the square of the GS sign information is expected to be absent[9]. Berry phases are also related to non-adiabatic coupling terms (NACTs) [10-12] and although these are accessible through linear-response time-dependent DFT (LR-TD-DFT)[13] the present paper is concerned of "static" DFT.

We consider 2-fold degeneracies and real Hamiltonians (no magnetic interactions). A basic notion is a 2D manifold of arbitrary parameters, $X$ and $Y$, that affect the external potential of a particle system (system I). The external potential is a function on the manifold $v(\mathbf{r};X,Y)$ and by solution of the Schrödinger equation, this potential produces a manifold of GS densities $n(\mathbf{r};X,Y)$. From the non-crossing rule it follows that in most 2D manifolds 2-fold degeneracies will appear as isolated points and higher degeneracies are practically never seen.[7, 14] A 2-fold degeneracy point can be assumed at the origin and polar coordinates used: $X = R\cos\phi$, $Y = R\sin\phi$. At any point near the origin degenerate perturbation theory shows that the ground and first excited eigenstates are orthogonal linear combinations of two degenerate orthonormal eigenstates $\tilde{\Psi}_j$ $(j=1,2)$ at the origin:

$$\begin{aligned}|\Psi_1(R,\phi)\rangle &= \cos\theta|\tilde{\Psi}_1\rangle + \sin\theta|\tilde{\Psi}_2\rangle \\ |\Psi_2(R,\phi)\rangle &= \sin\theta|\tilde{\Psi}_1\rangle - \cos\theta|\tilde{\Psi}_2\rangle\end{aligned} \quad (1)$$

with mixing angle $\theta$ a function of $R$ and $\phi$. We can speak of the $\phi$ dependent limit: $\theta(\phi) = \lim_{R\to 0}\theta(R,\phi)$. The density on the manifold $n(\mathbf{r};R,\phi)$ at point $R,\phi$ is easily calculated from Eq. (1) and the limit density $n(\mathbf{r};\phi) = \lim_{R\to 0} n(\mathbf{r};R,\phi)$ is:

$$n(\mathbf{r};\phi) = n_+(\mathbf{r}) + n_-(\mathbf{r})\cos 2\theta(\phi) + n_{12}(\mathbf{r})\sin 2\theta(\phi) \quad (2)$$

Where $2n_\pm = n_{11} \pm n_{22}$, $n_{ij}(\mathbf{r}) = \langle\tilde{\Psi}_i|\hat{n}(\mathbf{r})|\tilde{\Psi}_j\rangle$ and $\hat{n}(\mathbf{r})$ is the particle density operator. Eq. (2) shows that the limit density depends on $\phi$: the density we see when going into the degeneracy in one direction is different from that going in from another direction (Figure 1).

A compact way to characterize the scar is by considering the derivative with respect to $\phi$ $n'(\mathbf{r};\phi)$: a scar is present if it is not identically zero ($n'(\mathbf{r};\phi) \not\equiv 0$). We showed above that a degeneracy imposes a scar.

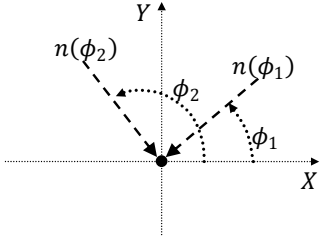

Figure 1: The "topological scar": $n(\phi_1)$ and $n(\phi_2)$ are particle densities along two distinct paths converging into the same manifold point. If they are different a GS degeneracy exists in system I.

Having discussed the density $n(\mathbf{r};R,\phi)$ in system I, which is PVR, we now consider this same density but impose it on some new system of particles with different particle-particle interaction. We call this system II. By the Hohenberg-Kohn theorem there is a unique a potential reproducing it, either as a pure-state or an ensemble density. However, since this density carries the scar at the origin the potentials there must reflect some irregularity. If the potential itself is unscarred (i.e. $v'_{II}(\mathbf{r};R\to 0,\phi) \equiv 0$) then the underlying wave function in system II must be degenerate, in such a way as to reproduce the density scar imposed by system I. However, the potential $v_{II}(\mathbf{r};R,\phi)$ in system II, which reproduces the scarred density, may itself be "scarred" (for example, when the density is scarred, the local density approximation for exchange potential $n(\mathbf{r};R,\phi)^{1/3}$ is scarred as well). In this case, the KS system might not develop a degeneracy at the origin even though the scar is reproduced.

What about the converse? Suppose the density is not scarred at the origin; can system II develop a degeneracy there? If $\tilde\Psi_k$, $k=1,\ldots,g$ are the degenerate states, then the first $g$ eigenstates in the immediate neighborhood, at direction $\phi$ are $\Psi_j(\phi) = \sum_{k=1}^{g} O_{jk}(\phi)\tilde\Psi_k$, where $O_{jk}$ are the elements of an orthogonal matrix. In the absence of a scar, the limit density $n(\mathbf{r};\phi) = \sum_{j=1}^{g} w_j(\phi)\langle\Psi_j(\phi)|\hat n(\mathbf{r})|\Psi_j(\phi)\rangle$ has to be independent of $\phi$, leading to the requirement that $\sum_{j=1}^{g} w_k O_{jk}(\phi) O_{j'k}(\phi)$ is independent of $\phi$. Unless special circumstances prevail, this can happen only if $w_k$ are all equal, i.e. $w_k = \frac{1}{g}$. We conclude that if the density is unscarred at the origin then either it is PVR, in which case there is no degeneracy, or it is an *equal-weight ensemble* (EWE) of *all* $g$ degenerate states.

Let now consider what one can deduce from a KS-DFT calculation pertaining to an external potential (such as that resulting from the nuclei in a molecule), assuming the exact XC potential is accessible. As one varies the potential parameters, the KS potential changes and the density of the physical system is reproduced, including the positions of its scars, signaling degeneracies in the physical system. As explained above, a corresponding degeneracy in the KS system is not mandatory, unless the KS potential itself is "unscarred". An absence of a density scar implies non-degeneracy in both systems. In the KS system this latter assertion rises because one cannot smoothly move from $g$-fold into $g'$-fold EWEs so the level of degeneracy on the manifold is "locked".

The next question is whether the Berry phase in the physical system traversing a loop around the scar can be reproduced by the exact KS calculation. The Berry phase[8] $\beta$ is equal to $\pi$ (0) if the real GS $\Psi_1(R,\phi)$ changes (does not change) its sign when carried smoothly once around the loop. If $\beta = \pi$ (0) the degeneracy is a "Jahn-Teller" or *conical* ("Renner-Teller") intersection. [10, 11, 15] The Berry phase can be calculated as an integral over the NACTs[10, 16]:

$$\beta = \lim_{R\to 0}\int_0^{2\pi} \langle\Psi_1(R,\phi)|\Psi_2'(R,\phi)\rangle d\phi, \quad (3)$$

and since from Eq. (1) $\langle\Psi_1|\Psi_2'\rangle = \theta'$, all we need for computing $\beta$ is to know $\theta'(\phi)$. We now show that this can be inferred directly from the density scar itself. Taking the third derivative of $n(\mathbf{r};\phi)$ in Eq. (2), we obtain:

$$n''' - 3gn'' = (-4\theta'^2 - 2g^2 + g')n', \quad (4)$$

where $g = \theta''/\theta'$. Using, for example, the x-y components of the dipole moment (DM) $\mathbf{d}(\phi) = \int n(\mathbf{r};\phi)\mathbf{r}\,d^3r$ we find after some manipulation:

$$\frac{d_x''' - 3gd_x''}{d_x'} = \frac{d_y''' - 3gd_y''}{d_y'} = -4\theta'^2 - 2g^2 + g', \quad (5)$$

From the first equality one obtains $g = \frac{1}{3}\frac{a'}{a}$ where $a = d_x'' d_y' - d_y'' d_x' = \mathbf{d}'' \times \mathbf{d}'$ and so:

$$\theta'(\phi) = B\,a(\phi)^{1/3}, \quad (6)$$

where $B$ is a constant (this constant can be determined by using the second equality in Eq.(5)). The value of $\theta(\phi) - \theta(0)$ can thus be determined by integrating Eq. (6) and in particular the Berry phase is recovered. This method of using using DM data is distinct from the Hush-Mulliken diabatization (HMD) (see [17] and references within). We use *only* GS DMs, while HMD uses the DM in *two* adiabatic states as well as the *transition* DM.

We demonstrate the validity of this method by a numerical example using the $H_3$ molecular system within an Extended Hückel (EH) approximation. Two fixed hydrogen nuclei are placed on the x-y plane at Cartesian points $\mathbf{R}_1 = \left(-3/2, \sqrt{3}/2, 0\right)a$, $\mathbf{R}_2 = -\left(3/2, \sqrt{3}/2, 0\right)a$, where $\sqrt{3}a = 1.4 a_0$ is the equilateral triangle edge length. Our 2D sub-manifold for the position of the third atom is the x-y plane and the $H_3$ CI is at the origin. The basis set for the EH calculation includes three Slater type 1s orbitals, one from each atom, $\chi_\alpha(\mathbf{r}) = 2\zeta^{3/2} e^{-\zeta|\mathbf{r}-\mathbf{R}_\alpha|}$, $\alpha = 1,2,3$ with exponent $\zeta = 1.3\, a_0^{-1}$. The Hamiltonian matrix $H_{EH}$ has diagonal elements $\left(H_{EH}\right)_{\alpha\alpha} = -0.5 E_h$ and off-diagonal elements $\left(E_{EH}\right)_{\alpha\beta} = KS_{\alpha\beta}$, $\alpha,\beta = 1,2,3$, where $S_{\alpha\beta} = \left\langle \chi_\alpha | \chi_\beta \right\rangle$ is the overlap matrix and $K = 0.875 E_h$. The generalized eigenvalue equation $H_{EH} Z = SZE$ is solved and the molecular orbital coefficients matrix $Z$ is used to construct the density matrix $P = ZpZ^T$ where $p$ is the diagonal matrix of occupation numbers (diag($p$) = (2,1,0) for three electrons in the $H_3$ doublet state). The DM in Cartesian direction $w = x, y$ is computed from $d_w = $ tr($D_w P$) where $\left(D_w\right)_{\alpha\beta} = e\left\langle \chi_\alpha | \mathbf{r}_w | \chi_\beta \right\rangle$ are the DM matrix elements in the basis set.

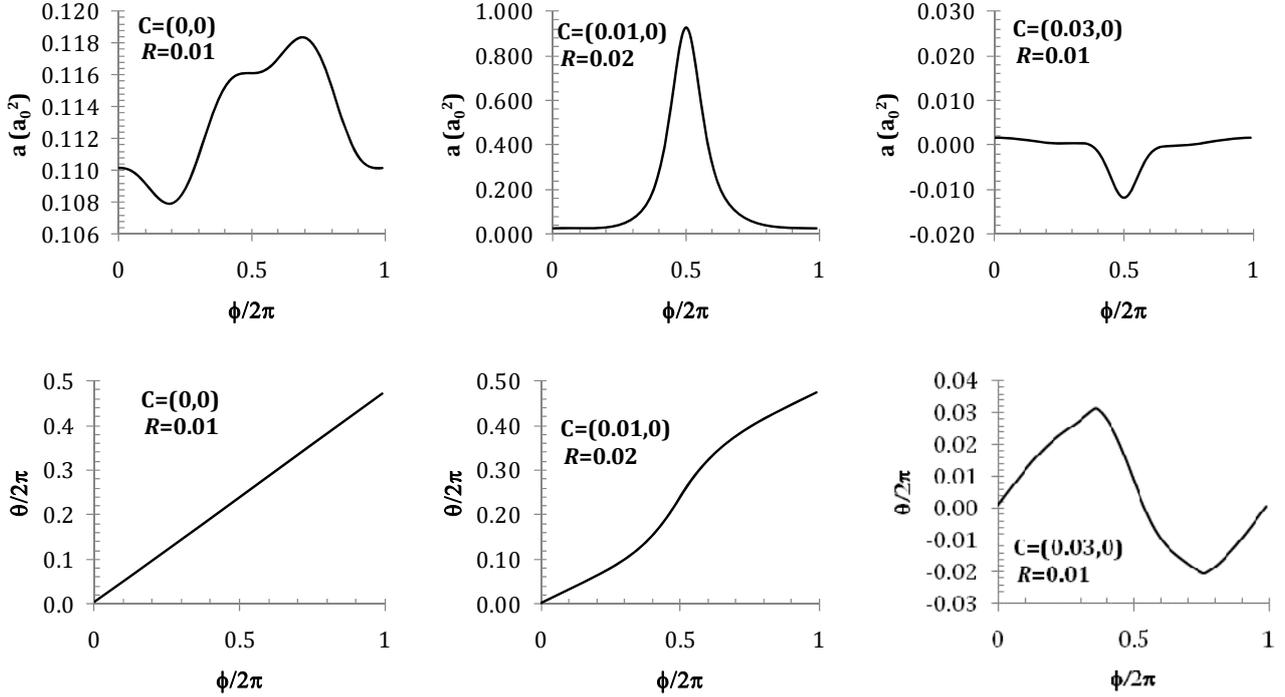

Figure 2: Reconstruction of the mixing angle of $H_3$ using the components of the electronic DM calculated along circular paths using the extended Hückel approximation. Three paths in the x-y plane are considered, each of radius $R\, a_0$ centered around the point $C$ as depicted in each panel. The CI is located at the origin. Top panels: The function $a(\phi)$. Bottom panels: the mixing angle $\theta(\phi)$.

We examine three cases where the third atom $\mathbf{R}_3$ moves along a circular loop encircling the point $C$ with radius $R$, shown in Figure 2. In the first two cases the loop encircles the origin so we expect the phase difference to be an odd multiple of $\pi$ while in the third case, not surrounding a CI, it should be 0. We calculated the GS DM components for $N_\phi = 100$ equally spaced angles in the range $[0, 2\pi]$ and from them, using discrete Fourier methods for calculating the required derivatives, we computed the function $a(\phi)$ and the mixing angle $\theta(\phi)$, both shown in Figure 2 for each case. For case 1 the radius $R$ is small enough for the theory to hold well and the final value of the mixing angle was $0.94\pi$. Case 2 shows a very different behavior of $a(\phi)$ but the final value of $\theta$ is still $0.95\pi$. The method is useful even when the loop is non concentric. In case 3 the path does not encircle the degeneracy and the mixing angle remains small along the path, its final value returning close to 0. It is difficult to enlarge $R$ in this method (accuracy quickly degrades $B$ changes) which breaks down when $\theta'^2$ acquires significantly negative values.

**Summary and Discussion**: We found that a pure-state density on some 2D parameter manifold produces topological scars at degeneracy points and only there. Such a scar is sufficiently detailed to allow reconstruction of the mixing angle and the Berry phase associated with the ground-state wavefunction of a small loop around the degeneracy. Any system of particles ("system II") that has this density as its ground state will thus "know" about the degeneracies in system I. Our results do not contradict the findings of ref.[6], namely that $g$-fold and $g'$-fold degeneracies have the same measure in density space. This is because our manifold makes a special cut through density space by considering densities that are known to be PVR in at least one system (system I) on the manifold of interest. While our findings show that exact KS-DFT can pinpoint and characterize degeneracies in the physi-cal system, any implementation of KS-DFT uses approximate functionals. In this latter case one cannot assume that the densities thus produced are PVR in any given system and there is no guarantee that the scars in the approximate KS system are points. Indeed, they could be 1D lines or even 2D regions of non-EWE densities (which we proved do not exist in the exact KS system). However, since non-EWEs cannot form without breaking symmetries, perhaps these problems do not easily arise at least in symmetrical molecules. The first step for developing DFT as a tool for studying degeneracies in molecules should therefore be benchmarking of the accuracy and reliability of various approximate functionals for locating and characterizing degeneracies in molecules.

We gratefully thank Dr. Nathan Argaman for invaluable insights to this manuscript. This work was supported by the Israel Science Foundation.